\documentclass[12pt]{article}
\usepackage{epsf}
\title{ {\bf Lepton flavor violating Z
boson decays induced by scalar unparticle}}
\author{\vspace{1cm}\\
        {\bf E. O. Iltan,}
        \thanks{E-mail address:
        eiltan@newton.physics.metu.edu.tr}
 \\
        Physics Department, Middle East Technical University \\
        Ankara, Turkey\\}

\date{}

\begin{document}
\setlength{\baselineskip}{24pt}
\maketitle
\setlength{\baselineskip}{7mm}
\begin{abstract}
We predict the branching ratios of the lepton flavor violating Z
boson decays $Z\rightarrow e^{\pm} \mu^{\pm}$, $Z\rightarrow
e^{\pm} \tau^{\pm}$ and $Z\rightarrow \mu^{\pm} \tau^{\pm}$  in
the case that the lepton flavor violation is carried by the scalar
unparticle mediation. We observe that their BRs are strongly
sensitive to the unparticle scaling dimension and  the branching
ratios can reach to the values of the order of $10^{-8}$, for the
heavy lepton flavor case, for the small values of the scaling
dimension.
\end{abstract}
\thispagestyle{empty}
\newpage
\setcounter{page}{1}
%%%
%%%
%
Lepton flavor violating (LFV) interactions reached great interest
since they are sensitive the physics beyond the standard model
(SM) and the related experimental measurements are improved at
present. $Z\rightarrow l_1\,l_2$ decays are among the LFV
interactions and the theoretical predictions of their branching
ratios (BRs) in the framework of the SM are extremely small
\cite{Riemann, Ganapathi, Illana}:
\begin{eqnarray}
BR(Z\rightarrow e^{\pm} \mu^{\pm})\sim BR(Z\rightarrow e^{\pm}
\tau^{\pm})
&\sim& 10^{-54}  \nonumber \, , \\
BR(Z\rightarrow \mu^{\pm} \tau^{\pm}) &<& 4\times 10^{-60}
\label{Theo1}
\end{eqnarray}
in the case of non-zero lepton mixing mechanism \cite{Pontecorvo}.
These results are far from the experimental limits obtained at
LEP1 \cite{PartData}:
\begin{eqnarray}
BR(Z\rightarrow e^{\pm} \mu^{\pm}) &<& 1.7\times 10^{-6} \,\,\,
\cite{Opal}
\nonumber \, , \\
BR(Z\rightarrow e^{\pm} \tau^{\pm}) &<& 9.8\times 10^{-6}\,\,\,
\cite{Opal,L3} \nonumber \, , \\
BR(Z\rightarrow \mu^{\pm} \tau^{\pm}) &<& 1.2\times 10^{-5} \,\,,
\cite{Opal,Delphi} \label{Expr1}
\end{eqnarray}
and from the improved ones at Giga-Z \cite{Wilson}:
\begin{eqnarray}
BR(Z\rightarrow e^{\pm} \mu^{\pm}) &<& 2\times 10^{-9}  \nonumber \, , \\
BR(Z\rightarrow e^{\pm} \tau^{\pm}) &<& f\times 6.5\times 10^{-8}
\nonumber \, , \\
BR(Z\rightarrow \mu^{\pm} \tau^{\pm}) &<& f\times 2.2\times
10^{-8} \label{Expr2}
\end{eqnarray}
with $f=0.2-1.0$\footnote{Notice that these numbers are obtained
for the decays $Z\rightarrow \bar{l}_1 l_2+ \bar{l}_2 l_1$ where
\begin{eqnarray}
BR(Z\rightarrow l_1^{\pm} l_2^{\pm})=\frac{\Gamma (Z\rightarrow
\bar{l}_1 l_2+ \bar{l}_2 l_1)}{\Gamma_Z} \nonumber .
\end{eqnarray}
}. On the other hand the Giga-Z option of the Tesla project aims
to increase the production of Z bosons at resonance
\cite{Hawkings}. These numerical values and the forthcoming
projects stimulate one to make theoretical works on the LFV Z
decays and to enhance their BRs by considering new scenarios
beyond the SM. There are various works related to these decays in
the literature \cite{Riemann}-\cite{Illana},
\cite{PartData}-\cite{Wilson}, \cite{Ghosal}-\cite{Yue}, namely
the extension of $\nu$SM with one (two) heavy ordinary Dirac
(right-handed singlet Majorana) neutrino(s) \cite{Illana}, the Zee
model \cite{Ghosal}, the  two Higgs doublet model (2HDM)
\cite{EiltZl1l2}, the 2HDM with extra dimensions
\cite{EiltZl1l2Extr, EiltZl1l2Split, EiltZl1l2LocNewHiggs}, the
supersymmetric models \cite{Masip, Cao}, top-color assisted
technicolor model \cite{Yue}.

In the present work, we consider that the lepton flavor (LF)
violation is carried by the scalar unparticle
($\textit{U}$)-lepton-lepton vertex and unparticles appear in the
internal line, in the loop. The unparticle idea is introduced by
Georgi \cite{Georgi1, Georgi2} and its effect in the processes,
which are induced at least in one loop level, is studied in
various works \cite{Lu}-\cite{IltanUn2}. The starting point of
this idea is the interaction of the SM and the ultraviolet sector,
having non-trivial infrared fixed point, at high energy level. The
ultraviolet sector comes out as new degrees of freedom, called
unparticles, being massless and having non integral scaling
dimension $d_u$ around, $\Lambda_U\sim 1\,TeV$. The effective
lagrangian which drives the interactions of unparticles with the
SM fields in the low energy level reads
\begin{equation}
{\cal{L}}_{eff}\sim
\frac{\eta}{\Lambda_U^{d_u+d_{SM}-n}}\,O_{SM}\, O_{U} \,,
\label{efflag}
\end{equation}
where $O_U$ is the unparticle operator, the parameter $\eta$ is
related to the energy scale of ultraviolet sector, the low energy
one and the matching coefficient \cite{Georgi1,Georgi2,Zwicky} and
$n$ is the space-time dimension.

\begin{figure}[htb]
\vskip 2.2truein \centering \epsfxsize=6.5in
\leavevmode\epsffile{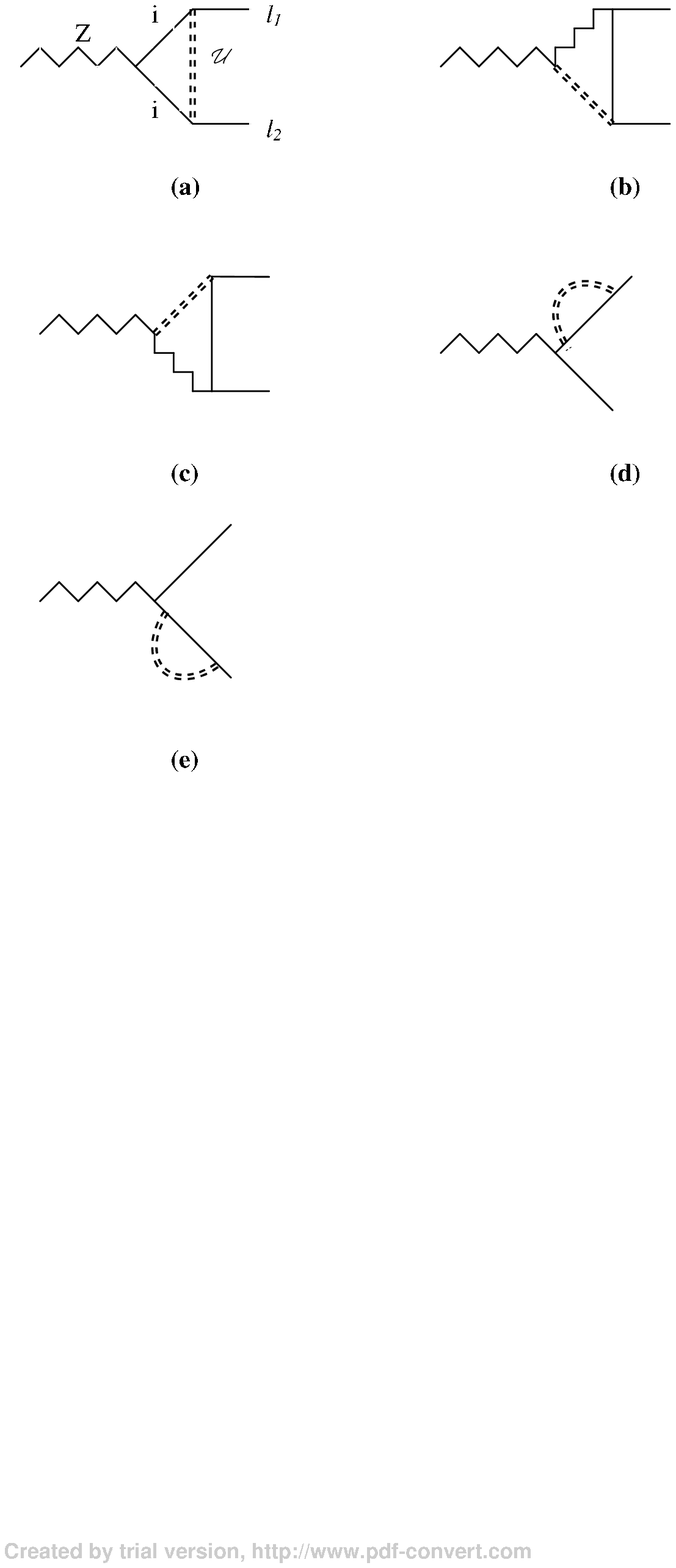} \vskip -5.5truein
\caption[]{One loop diagrams contribute to $Z\rightarrow
l_1^-\,l_2^+$ decay with scalar unparticle mediator. Solid line
represents the lepton field: $i$ represents the internal lepton,
$l_1^-$ ($l_2^+$) outgoing lepton (anti lepton), wavy line the Z
boson field, double dashed line the unparticle field.}
\label{figselfvert}
\end{figure}

At this stage, we choose the appropriate operators in order to
drive the LFV decays\footnote{Notice that the operators with the
lowest possible dimension are chosen since they have the most
powerful effect in the low energy effective theory (see for
example \cite{SChen}).}. The effective interaction lagrangian
responsible for the LFV decays in the low energy effective theory
is
\begin{eqnarray}
{\cal{L}}_1= \frac{1}{\Lambda_U^{du-1}}\Big (\lambda_{ij}^{S}\,
\bar{l}_{i} \,l_{j}+\lambda_{ij}^{P}\,\bar{l}_{i}
\,i\gamma_5\,l_{j}\Big)\, O_{U} \, , \label{lagrangianscalar}
\end{eqnarray}
where $l$ is the lepton field and $\lambda_{ij}^{S}$
($\lambda_{ij}^{P}$) is the scalar (pseudoscalar)  coupling. On
the other hand, there is a possibility that tree level
$\textit{U}-Z-Z$ interaction exists\footnote{The vertex factor:
$\frac{4\,i}{\Lambda_U^{d_u}}\,\lambda_0\,
(k_{1\nu}\,k_{2\mu}-k_1.k_2\,g_{\mu\,\nu})$
%+
%\lambda_0^{\prime}\,\epsilon_{\alpha\beta\mu\nu}\,k_1^\alpha\,k_2^\beta
%\Bigg)$
where $k_{1(2)}$ is the four momentum of Z boson with polarization
vector $\epsilon_{1\,\mu \,(2\,\nu)}$.} and it has a contribution
to the  LFV Z decays (see Fig \ref{figselfvert} (b) and (c)). The
corresponding effective Lagrangian reads
\begin{eqnarray}
{\cal{L}}_2= \frac{\lambda_0}{\Lambda_U^{du}}\,
F_{\mu\nu}\,F^{\mu\nu}\, O_{U}
%+\lambda_0^{\prime}\,
%\tilde{F}_{\mu\nu}\,F^{\mu\nu} \Big)\, O_{U}
\, , \label{lagrangianZ}
\end{eqnarray}
where $F_{\mu\nu}$ is the field tensor for the $Z_{\mu}$ field and
$\lambda_0$ is the effective coupling constant.
%$\tilde{F}_{\mu\nu}=\frac{1}{2}\epsilon_{\mu\nu\alpha\beta}\,
%F^{\alpha\beta}$.

The one loop  level $Z\rightarrow l_1\,l_2 $ decay (see
Fig.\ref{figselfvert}) is carried with the help of the scalar
unparticle propagator, which is obtained by using the scale
invariance \cite{Georgi2, Cheung1}:
\begin{eqnarray}
\!\!\! \int\,d^4x\,
e^{ipx}\,<0|T\Big(O_U(x)\,O_U(0)\Big)0>=i\frac{A_{d_u}}{2\,\pi}\,
\int_0^{\infty}\,ds\,\frac{s^{d_u-2}}{p^2-s+i\epsilon}=i\,\frac{A_{d_u}}
{2\,sin\,(d_u\pi)}\,(-p^2-i\epsilon)^{d_u-2} , \label{propagator}
\end{eqnarray}
with the factor $A_{d_u}$
\begin{eqnarray}
A_{d_u}=\frac{16\,\pi^{5/2}}{(2\,\pi)^{2\,d_u}}\,
\frac{\Gamma(d_u+\frac{1}{2})} {\Gamma(d_u-1)\,\Gamma(2\,d_u)} \,
. \label{Adu}
\end{eqnarray}
The function $\frac{1}{(-p^2-i\epsilon)^{2-d_u}}$ in eq.
(\ref{propagator}) becomes
\begin{eqnarray}
\frac{1}{(-p^2-i\epsilon)^{2-d_u}}\rightarrow
\frac{e^{-i\,d_u\,\pi}}{(p^2)^{2-d_u}} \, , \label{strongphase}
\end{eqnarray}
for $p^2>0$ and a non-trivial phase appears as a result of
non-integral scaling dimension.

\begin{figure}[htb]
\vskip -3.0truein \centering \epsfxsize=6.8in
\leavevmode\epsffile{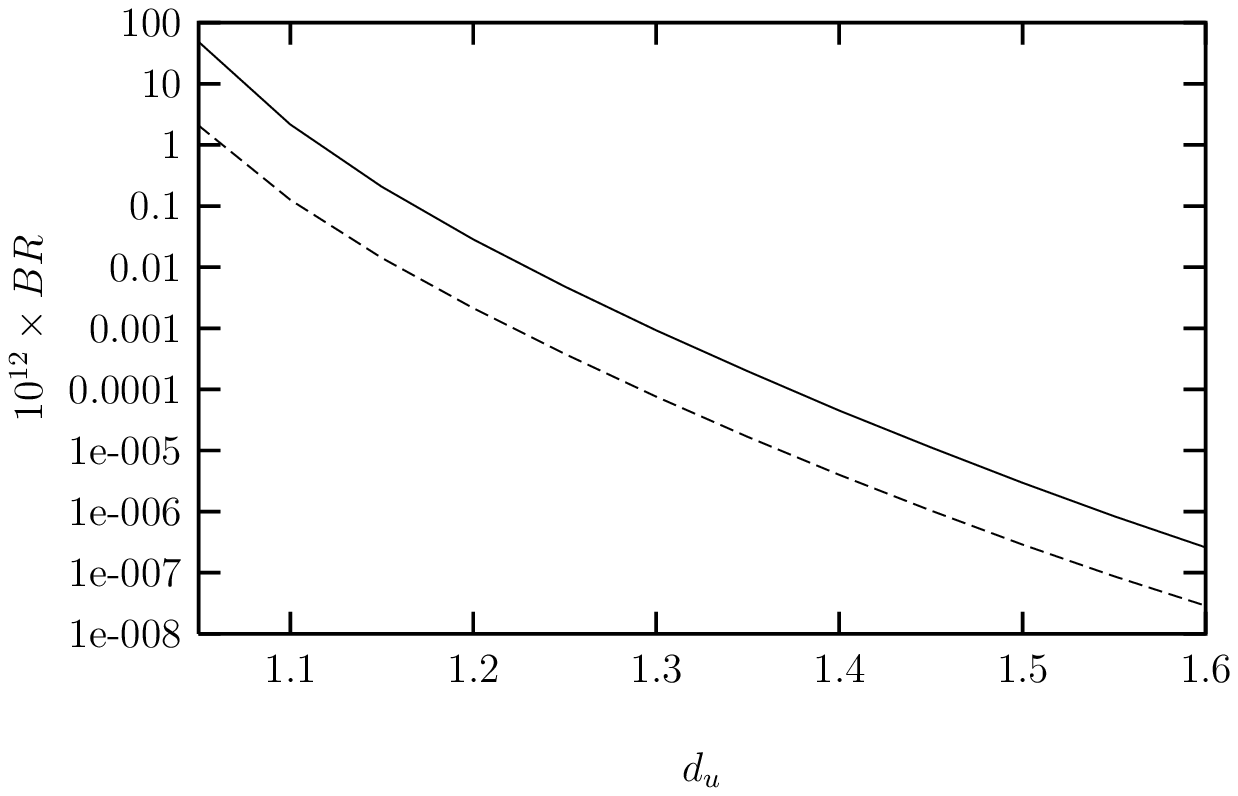} \vskip -3.0truein \caption[]{ The
scale parameter $d_u$ dependence of the BR $(Z\rightarrow
\mu^{\pm}\, e^{\pm})$ for $\Lambda_u=10\, TeV$, the couplings
$\lambda_{ee}=0.01$, $\lambda_{\mu\mu}=0.1$,
$\lambda_{\tau\tau}=1$ and $\lambda_{ij}=0.005$, $i\neq j$. Here
the solid (dashed) line represents the BR  for total contribution
and $\lambda_0=0.1$ (the contribution due to the $\textit{U}-Z-Z$
vertex and $\lambda_0=1$).} \label{Zmuedu}
\end{figure}

\begin{figure}[htb]
\vskip -3.0truein \centering \epsfxsize=6.8in
\leavevmode\epsffile{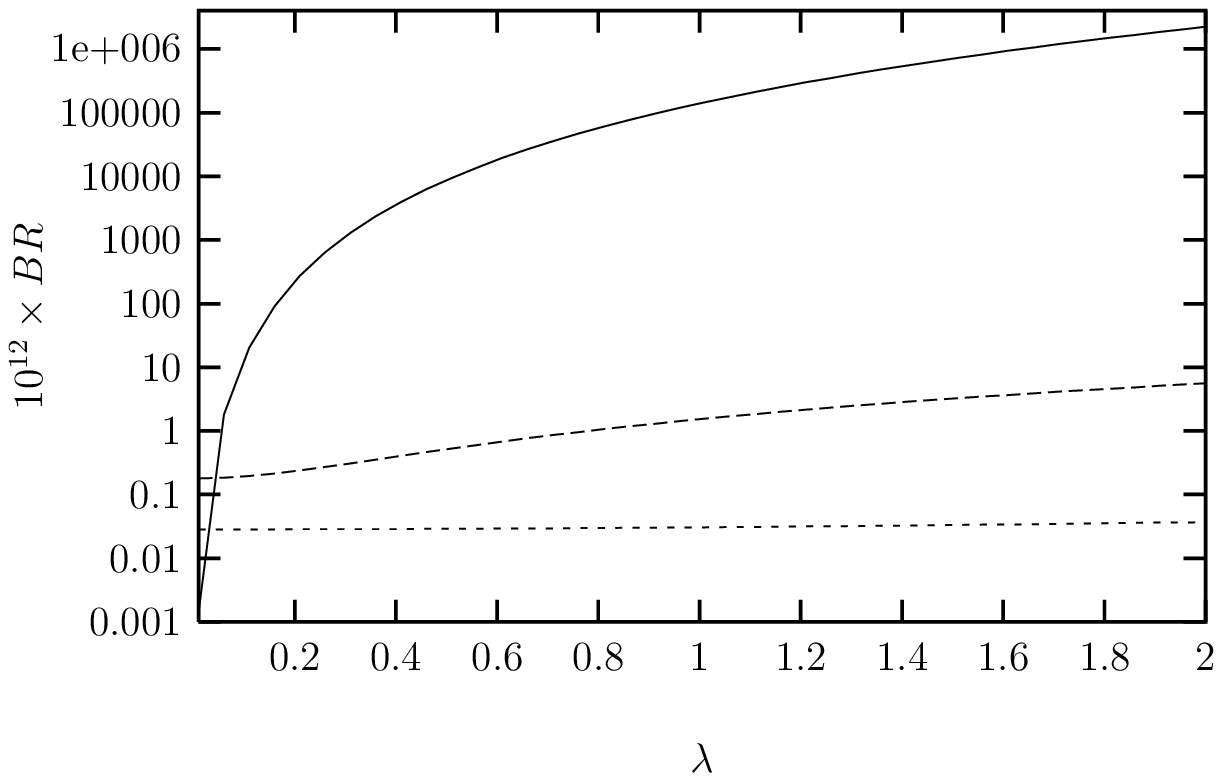} \vskip -3.0truein \caption[]{ The
BR $(Z\rightarrow \mu^{\pm}\, e^{\pm})$ with respect to the
couplings $\lambda$, for $d_u=1.2$. Here the solid (dashed-small
dashed) line represents the BR with respect to $\lambda$, for
$\lambda=\lambda_{ee}=\lambda_{\mu\mu}=\lambda_{\tau\tau}$,
$\lambda_{ij}=0.5\, \lambda$, $\lambda_0=0.1$ and $\Lambda_u=10\,
TeV$ (with respect to $\lambda_0$ for $\lambda_{ee}=0.01$,
$\lambda_{\mu\mu}=0.1$, $\lambda_{\tau\tau}=1$,
$\lambda_{ij}=0.005$, $\Lambda_u=1.0\, TeV$- with respect to
$\lambda_0$ for $\lambda_{ee}=0.01$, $\lambda_{\mu\mu}=0.1$,
$\lambda_{\tau\tau}=1$, $\lambda_{ij}=0.005$, $\Lambda_u=10\,
TeV$).} \label{Zmuelam}
\end{figure}

Now, we present the general effective vertex for the interaction
of on-shell Z-boson with a fermionic current:
\begin{eqnarray}
\Gamma_{\mu}=\gamma_{\mu}(f_V-f_A\ \gamma_5)+
\frac{i}{m_W}\,(f_M+f_E\, \gamma_5)\, \sigma_{\mu\,\nu}\, q^{\nu}
\, , \label{vertex}
\end{eqnarray}
where $q$ is the momentum transfer, $q^2=(p-p')^2$, $f_V$ ($f_A$)
is vector (axial-vector) coupling, $f_M$ ($f_E$) magnetic
(electric) transitions of unlike fermions. Here $p$
($-p^{\prime}$) is the four momentum vector of lepton
(anti-lepton). The form factors $f_V$, $f_A$, $f_{M}$ and $f_{E}$
in eq. (\ref{vertex}) are obtained as
\begin{eqnarray}
f_V&=&\int^{1}_{0}\,dx\,f_{V\,self}+\int^{1}_{0}\,dx\,\int^{1-x}_{0}\,dy\,
f_{V\,vert} \, ,\nonumber \\
f_A&=&\int^{1}_{0}\,dx\,f_{A\,self}+\int^{1}_{0}\,dx\,\int^{1-x}_{0}\,dy\,
f_{A\,vert} \, ,\nonumber \\
f_M&=&\int^{1}_{0}\,dx\,\int^{1-x}_{0}\,dy\, f_{M\,vert} \, ,\nonumber \\
f_E&=&\int^{1}_{0}\,dx\,\int^{1-x}_{0}\,dy\,f_{E\,vert} \, .
\label{funpart}
\end{eqnarray}
Taking into account all the masses of internal leptons and
external lepton (anti-lepton), the explicit expressions of
$f_{V\,self}$, $f_{A\,self}$, $f_{V\,vert}$, $f_{A\,vert}$,
$f_{M\,vert}$ and $f_{E\,vert}$ read as
\begin{eqnarray}
f_{V\,self}&=&
\frac{c_{self}\,(1-x)^{1-d_u}}{32\,s_W\,c_W\,\pi^2\, \Big(
m^2_{l_2^+}-m^2_{l_1^-}\Big)\,(1-d_u)}\, \sum_{i=1}^3\,\Bigg \{ \,
(L_{self}^{d_u-1}-L_{self}^{\prime d_u-1})\,\Bigg ( m_i\, \Big(
\nonumber \\ &\times& \!\!\! \!\!
(m_{l_1^-}\,c_2+m_{l_2^+}\,c_1)\, (\lambda_{il_1}^S+i\,
\lambda_{il_1}^P)\,(\lambda_{il_2}^P-i\,\lambda_{il_2}^S)-(m_{l_1^-}\,c_1+
m_{l_2^+}\,c_2)\,
(\lambda_{il_1}^S-i\,\lambda_{il_1}^P)\,(\lambda_{il_2}^P+i\,\lambda_{il_2}^S
)\, \Big) \nonumber \\&-& m_{l_1^-}\,m_{l_2^+}\,(1-x)\, \Big(
c_1\,(i\,\lambda_{il_1}^P-\,\lambda_{il_1}^S)\,(\lambda_{il_2}^P-i\,
\lambda_{il_2}^S)+
c_2\,(i\,\lambda_{il_1}^P+\lambda_{il_1}^S)\,(\lambda_{il_2}^P+i\,
\lambda_{il_2}^S)\Big) \Bigg)\nonumber \\&+&
(m^2_{l_2^+}\,L_{self}^{\prime
d_u-1}-m^2_{l_1^-}\,L_{self}^{d_u-1})\,(1-x)\,\Big(
c_1\,(i\,\lambda_{il_1}^P+\lambda_{il_1}^S)\,(\lambda_{il_2}^P+i\,
\lambda_{il_2}^S)\nonumber \\&+&
c_2\,(\lambda_{il_1}^P+i\,\lambda_{il_1}^S)\,(\lambda_{il_2}^S+i\,
\lambda_{il_2}^P) \Big) \Bigg\}\nonumber \, ,
\end{eqnarray}
\begin{eqnarray}
f_{A\,self}&=&
\frac{c_{self}\,(1-x)^{1-d_u}}{32\,s_W\,c_W\,\pi^2\, \Big(
m_{l_2^+}-m_{l_1^-}\Big)\,(1-d_u)}\, \sum_{i=1}^3 \, \Bigg\{
(L_{self}^{d_u-1}-L_{self}^{\prime d_u-1})\,\Bigg ( m_i\,\Big(
\nonumber \\ &\times& \!\!\! \!\!
(m_{l_2^+}\,c_1-m_{l_1^-}\,c_2)\,
(\lambda_{il_1}^S+i\,\lambda_{il_1}^P)\,(\lambda_{il_2}^P-i\,\lambda_{il_2}^S)
-(m_{l_1^-}\,c_1-m_{l_2^+}\,c_2)
(\lambda_{il_1}^S-i\,\lambda_{il_1}^P)\,(\lambda_{il_2}^P+i\,\lambda_{il_2}^S)
\, \Big)\nonumber \\ &-& m_{l_1^-}\,m_{l_2^+}\,(1-x)\, \Big(
c_1\,(\lambda_{il_1}^P+i\,\lambda_{il_1}^S)\,(i\,\lambda_{il_2}^P+
\lambda_{il_2}^S)+
c_2\,(\lambda_{il_1}^P-i\,\lambda_{il_1}^S)\,(\lambda_{il_2}^S-i\,
\lambda_{il_2}^P)\Big)\Bigg)\nonumber \\
&+& (m^2_{l_2^+}\,L_{self}^{\prime
d_u-1}-m^2_{l_1^-}\,L_{self}^{d_u-1})\,(1-x)\,\Big(
c_1\,(\lambda_{il_1}^S+i\,\lambda_{il_1}^P)\,(\lambda_{il_2}^P+i\,
\lambda_{il_2}^S)\nonumber \\
&+&c_2\,(\lambda_{il_1}^S-i\,\lambda_{il_1}^P)\,
(\lambda_{il_2}^P-i\,\lambda_{il_2}^S) \Big)\Bigg\} \nonumber \, ,
\end{eqnarray}
\begin{eqnarray}
f_{V\,vert}&=& \frac{-c_{ver}\,(1-x-y)^{1-d_u}}{32\,\pi^2}\,
\sum_{i=1}^3\,\frac{1}{\,L_{vert}^{2-d_u}}\,
 \Bigg\{m_i\, (1-x-y\,)\nonumber \\  &\times& \!\!\!\!\!
 \Big( (m_{l_2^+}\,c_1+m_{l_1^-}\,c_2)\,
(\lambda_{il_1}^S+i\,\lambda_{il_1}^P)\,(\lambda_{il_2}^P-i\,\lambda_{il_2}^S)
-(m_{l_1^-}\,c_1+m_{l_2^+}\,c_2)\,
\,(\lambda_{il_1}^S-i\,\lambda_{il_1}^P)\,(\lambda_{il_2}^P+i\,
\lambda_{il_2}^S) \Big) \nonumber \\&-& m_i^2\,\Big(
c_1\,(\lambda_{il_1}^S+i\,\lambda_{il_1}^P)\,(\lambda_{il_2}^P+i\,
\lambda_{il_2}^S)+ \,c_2\,
(i\,\lambda_{il_1}^P-\lambda_{il_1}^S)\,(\lambda_{il_2}^P-i\,
\lambda_{il_2}^S)\Big)\nonumber \\&-&
\Big(c_1\,(i\,\lambda_{il_1}^P-\lambda_{il_1}^S)\,
(\lambda_{il_2}^P-i\,\lambda_{il_2}^S) + c_2\,
(i\,\lambda_{il_1}^P+\lambda_{il_1}^S)\,(\lambda_{il_2}^P+i\,
\lambda_{il_2}^S)\Big)\nonumber \\ &\times& \Big(
m^2_Z\,x\,y+m_{l_1^-}\,m_{l_2^+}\,(1-x-y)^2-
\frac{L_{vert}}{1-d_u} \Big) \Bigg\}
\nonumber \\ &-& \frac{\lambda_0\, m_Z^2}{16\,\pi^2}\,
\sum_{i=1}^3\,\Bigg\{
\frac{b_{ver}\,y^{1-d_u}}{L_{1\,vert}^{2-d_u}}\, \Big\{
(\lambda_{il_2}^P+i\,\lambda_{il_2}^S)\,\Bigg(c_1\,m_i\,(1-x+y)+c_2\,
 \Big(m_{l_1^-}\,x\,(x+y-1) \nonumber \\ &+&
m_{l_2^+}\,y\,(1+x+y) \Big)\Bigg) -
(\lambda_{il_2}^P-i\,\lambda_{il_2}^S)\,\Bigg( c_2\,m_i\,(1-x+y)+
 c_1\,\Big(m_{l_1^-}\,x\,(x+y-1)\nonumber \\ &+& \!\!\!
m_{l_2^+}\,y\,(1+x+y) \Big)\Bigg)\Big\}-
\frac{b^\prime_{ver}\,x^{1-d_u}}{L_{2\,vert}^{2-d_u}}\,\Big\{
(\lambda_{il_1}^P-i\,\lambda_{il_1}^S)\, \Bigg(
c_1\,m_i\,(1+x-y)+c_2\, \Big(m_{l_2^+}\,y\,(x+y-1) \nonumber \\
&+& m_{l_1^-}\,x\,(1+x+y) \Big)\Bigg)-
(\lambda_{il_1}^P+i\,\lambda_{il_1}^S)\,\Bigg( c_2\,m_i\,(1+x-y)+
 c_1\,\Big(m_{l_2^+}\,y\,(x+y-1)\nonumber \\ &+& m_{l_1^-}\,x\,(1+x+y)
 \Big)\Bigg)\Big\} \Bigg\} \nonumber \, ,
\end{eqnarray}
\begin{eqnarray}
f_{A\,vert}&=& \frac{-c_{ver}\,(1-x-y)^{1-d_u}}{32\,\pi^2}\,
\sum_{i=1}^3\,\frac{1}{\,L_{vert}^{2-d_u}}\,
 \Bigg\{ m_i\, (1-x-y\,)\nonumber \\ &\times& \!\!\!\!\!
 \Big( (m_{l_2^+}\,c_1-m_{l_1^-}\,c_2)\,
(\lambda_{il_1}^S+i\,\lambda_{il_1}^P)\,(\lambda_{il_2}^P-i\,\lambda_{il_2}^S)
-(m_{l_2^+}\,c_2-m_{l_1^-}\,c_1)\,
\,(i\,\lambda_{il_1}^P-\lambda_{il_1}^S)\,(\lambda_{il_2}^P+i\,
\lambda_{il_2}^S)\Big)\nonumber \\ &+& m_i^2\,\Big(
c_1\,(\lambda_{il_1}^P-i\,\lambda_{il_1}^S)\,(\lambda_{il_2}^S-i\,
\lambda_{il_2}^P)+\,c_2\,
(\lambda_{il_1}^P+i\,\lambda_{il_1}^S)\,(\lambda_{il_2}^S+i\,\lambda_{il_2}^P)
\Big) +\Big(c_1\,(\lambda_{il_1}^P+i\,\lambda_{il_1}^S) \nonumber \\
&\times&  (\lambda_{il_2}^S+i\,\lambda_{il_2}^P) + \,c_2\,
(\lambda_{il_1}^P-i\,\lambda_{il_1}^S)\,(\lambda_{il_2}^S-i\,
\lambda_{il_2}^P)\Big)\,
\Big(m^2_Z\,x\,y-m_{l_1^-}\,m_{l_2^+}\,(1-x-y)^2-
\frac{L_{vert}}{1-d_u} \Big) \Bigg\} \nonumber \\ &+&
\frac{\lambda_0\, m_Z^2}{16\,\pi^2}\, \sum_{i=1}^3\,\Bigg\{
\frac{b_{ver}\,y^{1-d_u}}{L_{1\,vert}^{2-d_u}}\, \Big\{
(\lambda_{il_2}^P+i\,\lambda_{il_2}^S)\,\Bigg(c_1\,m_i\,(x-y-1)+c_2\,
 \Big(m_{l_1^-}\,x\,(1-x-y) \nonumber \\ &+&
m_{l_2^+}\,y\,(1+x+y) \Big)\Bigg)+
(\lambda_{il_2}^P-i\,\lambda_{il_2}^S)\,\Bigg( c_2\,m_i\,(x-y-1)+
 c_1\,\Big(m_{l_1^-}\,x\,(1-x-y)\nonumber \\ &+& \!\!\!
m_{l_2^+}\,y\,(1+x+y) \Big)\Bigg)\Big\} +
\frac{b^\prime_{ver}\,x^{1-d_u}}{\,L_{2\,vert}^{2-d_u}}\,\Big\{
(\lambda_{il_1}^P-i\,\lambda_{il_1}^S)\, \Bigg(
c_1\,m_i\,(1+x-y)+c_2\, \Big(m_{l_2^+}\,y\,(x+y-1) \nonumber \\
&-& m_{l_1^-}\,x\,(1+x+y) \Big)\Bigg)+
(\lambda_{il_1}^P+i\,\lambda_{il_1}^S)\,\Bigg( c_2\,m_i\,(1+x-y)+
 c_1\,\Big(m_{l_2^+}\,y\,(x+y-1)\nonumber \\ &-& m_{l_1^-}\,y\,(1+x+y)
 \Big)\Bigg)\Big\} \Bigg\} \nonumber \, ,
\end{eqnarray}
\newpage
\begin{eqnarray}
f_{M\,vert}&=& \frac{-i\,(1-x-y)^{1-d_u}}{32\,\pi^2}\,
\sum_{i=1}^3\,\frac{c_{ver}\,m_Z\,c_W}{L_{vert}^{2-d_u}}\,
\Bigg\{m_i\,
 \Big((x+y)\,
(\lambda_{il_1}^S\,\lambda_{il_2}^S-\lambda_{il_1}^P\,\lambda_{il_2}^P)
(c_1+c_2)\nonumber \\&-& i\,(x-y)\,
(\lambda_{il_1}^S\,\lambda_{il_2}^P+\lambda_{il_1}^P\,\lambda_{il_2}^S)
(c_2-c_1)\Big)+(1-x-y)\,(m_{l_1^-}\,x+m_{l_2^+}\,y)\nonumber
\\ &\times&
%(m_{l_1^-}\,x-m_{l_2^+}\,y)
\,\Big(c_1\,
(\lambda_{il_1}^P+i\,\lambda_{il_1}^S)\,(\lambda_{il_2}^P-i\,\lambda_{il_2}^S)
+c_2\,(\lambda_{il_1}^P-i\,\lambda_{il_1}^S)\,(\lambda_{il_2}^P+i\,
\lambda_{il_2}^S)\, \Big)\Bigg \}
\nonumber \\ &-& \frac{i\,\lambda_0}{16\,\pi^2}\,
\sum_{i=1}^3\,\Bigg\{
\frac{b_{ver}\,m_Z\,c_W\,y^{1-d_u}}{L_{1\,vert}^{2-d_u}}\,\Bigg(\,\Big(
c_1\,(\lambda_{il_2}^S+i\,\lambda_{il_2}^P)+c_2\,
(\lambda_{il_2}^S-i\,\lambda_{il_2}^P)\Big)\nonumber \\
&\times& \Big(2\,m^2_Z\,x\,y+(1-x-y)\,\Big( m^2_{l_1^-
}\,x+m^2_{l_2^+}\,y- m_{l_1^-}\,m_{l_2^+}\,(x+y)\Big) -
2\,\frac{L_{1\,vert}}{1-d_u} \Big) \nonumber \\&-& \Big(
c_1\,(\lambda_{il_2}^S-i\,\lambda_{il_2}^P)+c_2\,
(\lambda_{il_2}^S+i\,\lambda_{il_2}^P)\Big)\,(1-x-y)\,
m_i\,(m_{l_1^-}- m_{l_2^+})\,\Bigg)
\nonumber \\ &+&
\frac{b^\prime_{ver}\,m_Z\,c_W\,x^{1-d_u}}{L_{2\,vert}^{2-d_u}}\,
\Bigg(\,\Big( c_1\,(\lambda_{il_1}^S-i\,\lambda_{il_1}^P)+c_2\,
(\lambda_{il_1}^S+i\,\lambda_{il_1}^P)\Big)\, \nonumber \\
&\times& \Big( 2\,m^2_Z\,x\,y+ (1-x-y)\,\Big(
m^2_{l_1^-}\,x+m^2_{l_2^+ }\,y- m_{l_1^-}\,m_{l_2^+}\,(x+y)\Big)-
2\,\frac{L_{2\,vert}}{1-d_u} \Big) \nonumber \\ &+& \Big(
c_1\,(\lambda_{il_1}^S+i\,\lambda_{il_1}^P)+c_2\,
(\lambda_{il_1}^S-i\,\lambda_{il_1}^P)\Big)\,(1-x-y)\,
m_i\,(m_{l_1^-}- m_{l_2^+})\Bigg) \Bigg\} \nonumber \, ,
\end{eqnarray}
\begin{eqnarray}
f_{E\,vert}&=& \frac{-i\,(1-x-y)^{1-d_u}}{32\,\pi^2}\,
\sum_{i=1}^3\,\frac{c_{ver}\,m_Z\,c_W}{L_{vert}^{2-d_u}}\,
\Bigg\{m_i\,
 \Big(i\,(x+y)\,
(\lambda_{il_1}^S\,\lambda_{il_2}^P+\lambda_{il_1}^P\,\lambda_{il_2}^S)\,
(c_1+c_2)\nonumber \\
&+& (x-y)\,
(\lambda_{il_1}^P\,\lambda_{il_2}^P-\lambda_{il_1}^S\,\lambda_{il_2}^S)\,
(c_2-c_1)\Big) +(1-x-y)\,(m_{l_1^-}\,x-m_{l_2^+}\,y)\nonumber
\\ &\times&
%(m_{l_1^-}\,x-m_{l_2^+}\,y)\,
\Big(c_1\,
(\lambda_{il_1}^P+i\,\lambda_{il_1}^S)\,(\lambda_{il_2}^P-i\,\lambda_{il_2}^S)
-c_2\,(\lambda_{il_1}^P-i\,\lambda_{il_1}^S)\,
(\lambda_{il_2}^P+i\,\lambda_{il_2}^S) \, \Big)\Bigg\}
\nonumber \\ &-& \frac{i\,\lambda_0}{16\,\pi^2}\,
\sum_{i=1}^3\,\Bigg\{
\frac{b_{ver}\,m_Z\,c_W\,y^{1-d_u}}{L_{1\,vert}^{2-d_u}}\,\Bigg(\,\Big(
c_1\,(\lambda_{il_2}^S+i\,\lambda_{il_2}^P)-c_2\,
(\lambda_{il_2}^S-i\,\lambda_{il_2}^P)\Big)\, \nonumber \\
&\times& \Big(2\,m^2_Z\,x\,y+(1-x-y)\,\Big(
m^2_{l_1^-}\,x+m^2_{l_2^+}\,y+
m_{l_1^-}\,m_{l_2^+}\,(x+y)\Big)-2\,\frac{L_{1\,vert}}{1-d_u}
\Big) \nonumber \\ &+& \Big(
c_1\,(\lambda_{il_2}^S-i\,\lambda_{il_2}^P)-c_2\,
(\lambda_{il_2}^S+i\,\lambda_{il_2}^P)\Big)\,(1-x-y)\,
m_i\,(m_{l_1^-}+ m_{l_2^+})\Bigg)
\nonumber \\ &-&
\frac{b^\prime_{ver}\,m_Z\,c_W\,x^{1-d_u}}{L_{1\,vert}^{2-d_u}}\,
\Bigg(\,\Big(c_1\,(\lambda_{il_1}^S-i\,\lambda_{il_1}^P)-c_2\,
(\lambda_{il_1}^S+i\,\lambda_{il_1}^P)\Big)\, \nonumber \\
&\times& \Big( 2\,m^2_Z\,x\,y+(1-x-y)\,\Big(
m_{l_1^-}\,x+m^2_{l_2^+}\,y+ m_{l_1^-}\,m_{l_2^+}\,(x+y)\Big)-
2\,\frac{L_{2\,vert}}{1-d_u} \Big) \nonumber \\
&+& \Big( c_1\,(\lambda_{il_1}^S+i\,\lambda_{il_1}^P)-c_2\,
(\lambda_{il_1}^S-i\,\lambda_{il_1}^P)\Big)\,(1-x-y)\,
m_i\,(m_{l_1^-}+ m_{l_2^+})\Bigg)  \Bigg\} \label{fAVME} \, ,
\end{eqnarray}
%%%%%%%%%
%
with
\begin{eqnarray}
L_{self}&=&x\,\Big(m_{l_1^-}^2\,(1-x)-m_i^2\Big)
\, , \nonumber \\
L_{self}^{\prime}&=&x\,\Big(m_{l_2^+}^2\,(1-x)-m_i^2\Big) \, ,
\nonumber \\
L_{vert}&=&(m_{l_1^-}^2\,x+m_{l_2^+}^2\,y)\,(1-x-y)-m_i^2\,(x+y)+m_Z^2\,x\,y
\, , \nonumber \\
L_{1\,vert}&=&
\Big(m_{l_1^-}^2\,x+m_{l_2^+}^2\,y-m_i^2\Big)\,(1-x-y)+m_Z^2\,x\,(y-1)
\, , \nonumber \\
L_{2\,vert}&=&
\Big(m_{l_1^-}^2\,x+m_{l_2^+}^2\,y-m_i^2\Big)\,(1-x-y)+m_Z^2\,y\,(x-1)
\label{Ll} \, ,
\end{eqnarray}
and
\begin{eqnarray}
c_{self}&=&-\frac{e\,A_{d_u}}{2\,sin\,(d_u\pi)\,\Lambda_u^{2\,(d_u-1)}}\,
, \nonumber \\
c_{ver}&=&-\frac{e\,A_{d_u}}{2\,s_W\,c_W\,\,sin\,(d_u\pi)\,\Lambda_u^{2\,(d_u-1)}}
\, , \nonumber \\
b_{ver}&=&-\frac{e\,A_{d_u}}{2\,s_W\,c_W\,\,sin\,(d_u\pi)\,\Lambda_u^{2\,d_u-1}}
\, , \nonumber \\
b^\prime_{ver}&=&-b_{ver} \label{cselfver} \, .
\end{eqnarray}
In eq. (\ref{fAVME}), the flavor changing scalar and pseudoscalar
couplings $\lambda_{il_{1(2)}}^{S,P}$ represent the effective
interaction between the internal lepton $i$, ($i=e,\mu,\tau$) and
the outgoing $l_1^-\,(l_2^+)$ lepton (anti lepton). Finally, the
BR for $Z\rightarrow l_1^-\,l_2^+$ can be obtained by using the
form factors $f_V$, $f_A$, $f_M$ and $f_E$ as
\begin{eqnarray}
BR (Z\rightarrow l_1^-\,l_2^+)=\frac{1}{48\,\pi}\,
\frac{m_Z}{\Gamma_Z}\,
\{|f_V|^2+|f_A|^2+\frac{1}{2\,cos^2\,\theta_W} (|f_M|^2+|f_E|^2)
\}  \label{BR1} \, ,
\end{eqnarray}
where
%$\alpha_W=\frac{g^2}{4\,\pi}$ and
$\Gamma_Z$ is the total
decay width of Z boson. Note that, in general, the production of
sum of charged states is considered with the corresponding BR
\begin{eqnarray}
BR (Z\rightarrow l_1^{\pm}\,l_2^{\pm})= \frac{\Gamma(Z\rightarrow
(\bar{l}_1\,l_2+\bar{l}_2\,l_1)}{\Gamma_Z} \, , \label{BR2}
\end{eqnarray}
and in our numerical analysis we use this branching ratio.
%
%%%
%%%
%\section{Discussion}
\\ \\
{\Large \textbf{Discussion}}
\\ \\
In this section, we estimate the BRs of LFV  Z boson decays by
considering that the flavor violation is carried by the scalar
unparticle mediation. These decays exist at least in one loop
level and, in the present case, we assume that the possible
sources of LF violation are the $\textit{U}$-lepton-lepton
couplings in the framework of the effective theory. On the other
hand, we take the $\textit{U}-Z-Z$ coupling non-zero and we study
the sensitivity of the BRs to this coupling. The couplings
considered and the scaling dimension of unparticle(s) are free
parameters and they should be restricted by respecting the current
experimental measurements and some theoretical considerations. For
the scaling dimension $d_u$ we choose the range\footnote{Here,
$d_u>1$ is due to the non-integrable singularities in the decay
rate \cite{Georgi2} and $d_u<2$ is due to the convergence of the
integrals \cite{Liao1}.} $1< d_u <2$. For the LF violating
couplings we consider the following restrictions:
\begin{itemize}
\item The (off) diagonal couplings are flavor (blind and
universal) aware and
$\lambda_{\tau\tau}>\lambda_{\mu\mu}>\lambda_{ee}$ ($\lambda_{ij},
i\neq j$).  We take the greatest numerical value of diagonal
coupling of the order of one and the off diagonal one as
$\lambda_{ij}=\kappa \lambda_{ee}$ with $\kappa < 1$. In our
numerical calculations, we choose $\kappa=0.5$.
\item As a second possibility, we consider that the (off) diagonal
couplings are flavor blind and universal and of the order of one.
Similar to the previous case, we take the off diagonal ones as
$\lambda_{ij}=\kappa \lambda_{ii}$ with $\kappa=0.5$.
\end{itemize}
Furthermore, we choose the coupling $\lambda_0$ for the tree level
$\textit{U}-Z-Z$ interaction (see eq. (\ref{lagrangianZ})) in the
range $0.1-1.0$ and we take the energy scale of the order of TeV.
Notice that throughout our calculations we use the input values
given in Table (\ref{input}).
\begin{table}[h]
        \begin{center}
        \begin{tabular}{|l|l|}
        \hline
        \multicolumn{1}{|c|}{Parameter} &
                \multicolumn{1}{|c|}{Value}     \\
        \hline \hline
        $m_e$           & $0.0005$   (GeV)  \\
        $m_{\mu}$                   & $0.106$ (GeV) \\
        $m_{\tau}$                  & $1.780$ (GeV) \\
        $\Gamma^{Tot}_Z$           & $2.49$ (GeV) \\
        $s_W^2$             & $0.23$  \\
        \hline
        \end{tabular}
        \end{center}
\caption{The values of the input parameters used in the numerical
          calculations.}
\label{input}
\end{table}
\\
In  Fig.\ref{Zmuedu}, we present the BR $(Z\rightarrow \mu^{\pm}\,
e^{\pm})$ with respect to the scale parameter $d_u$, for the
energy scale $\Lambda_u=10\, TeV$, the couplings
$\lambda_{ee}=0.01$, $\lambda_{\mu\mu}=0.1$,
$\lambda_{\tau\tau}=1$ and $\lambda_{ij}=0.005$, $i\neq j$. Here
the solid (dashed) line represents the BR  for total contribution
and $\lambda_0=0.1$ (the contribution due to the $\textit{U}-Z-Z$
vertex and $\lambda_0=1$). The BR is strongly sensitive to the
scale $d_u$ and, reaches to the numerical values $10^{-10}$, for
$d_u<1.1$. The contribution of $\textit{U}-Z-Z$ vertex is almost
two order smaller than the total one, even for $\lambda_0=1$. With
the increasing values of the scaling dimension $d_u$, the BR
sharply decreases and becomes negligible. Fig. \ref{Zmuelam}
represents the BR $(Z\rightarrow \mu^{\pm}\, e^{\pm})$ with
respect to the couplings $\lambda$, for $d_u=1.2$. Here the solid
(dashed-small dashed) line represents the BR with respect to
$\lambda$, for
$\lambda=\lambda_{ee}=\lambda_{\mu\mu}=\lambda_{\tau\tau}$,
$\lambda_{ij}=0.5\, \lambda$, $\lambda_0=0.1$ and $\Lambda_u=10\,
TeV$ (with respect to $\lambda_0$ for $\lambda_{ee}=0.01$,
$\lambda_{\mu\mu}=0.1$, $\lambda_{\tau\tau}=1$,
$\lambda_{ij}=0.005$, $\Lambda_u=1.0\, TeV$- with respect to
$\lambda_0$ for $\lambda_{ee}=0.01$, $\lambda_{\mu\mu}=0.1$,
$\lambda_{\tau\tau}=1$, $\lambda_{ij}=0.005$, $\Lambda_u=10\,
TeV$). In the case that the diagonal (off diagonal) couplings are
flavor blind, the BR can reach to the values of the order of
$10^{-7}$ for $\lambda=1$. This can ensure a valuable information
about the unparticle physics and the LFV couplings with more
accurate measurements of the decays under consideration.
Furthermore, this figure shows that the BR is not sensitive to the
coupling $\lambda_0$ and it enhances almost one order in the range
$0.1\leq \lambda_0\leq 1.0$, for the energy $\Lambda_u=1.0\, TeV$.

\begin{figure}[htb]
\vskip -3.0truein \centering \epsfxsize=6.8in
\leavevmode\epsffile{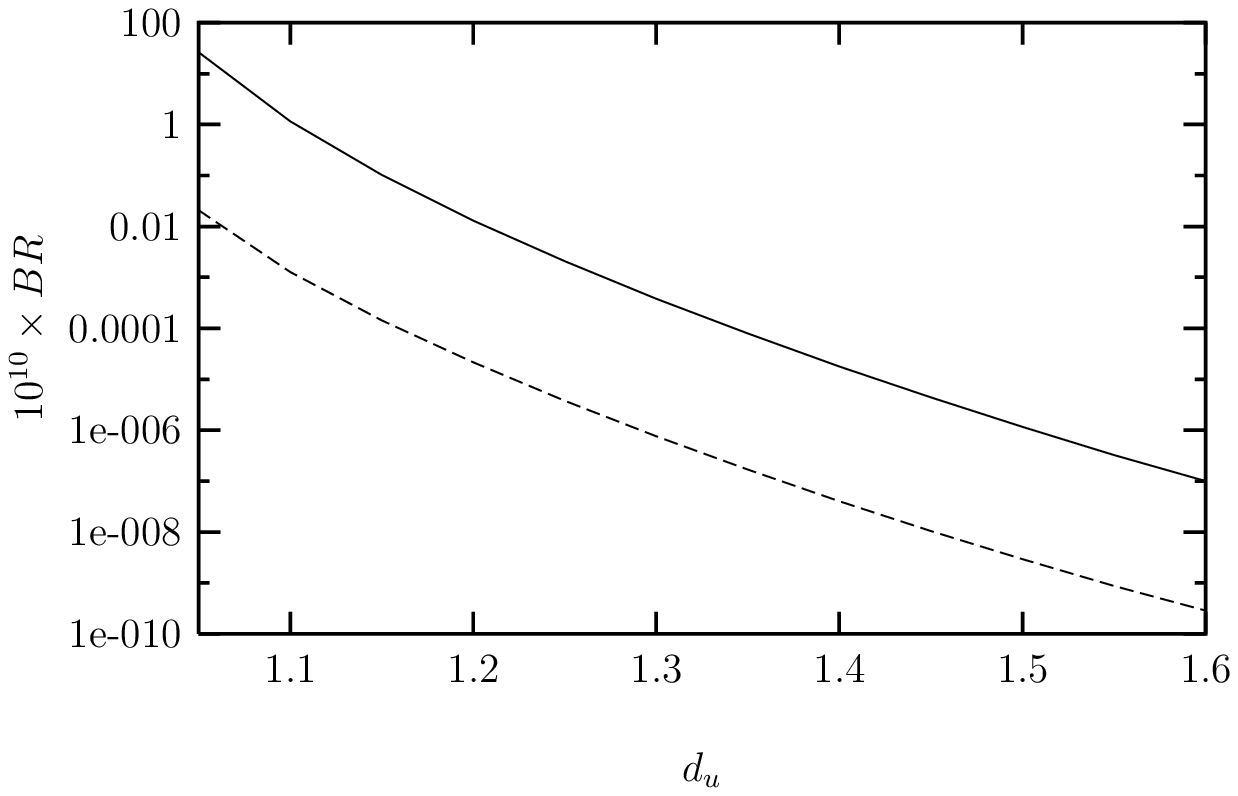} \vskip -3.0truein \caption[]{The
scale parameter $d_u$ dependence of the BR $(Z\rightarrow
\tau^{\pm}\, \mu^{\pm})$ for $\Lambda_u=10\, TeV$, the couplings
$\lambda_{ee}=0.01$, $\lambda_{\mu\mu}=0.1$,
$\lambda_{\tau\tau}=1$ and $\lambda_{ij}=0.005$, $i\neq j$. Here
the solid (dashed) line represents the BR  for total contribution
and $\lambda_0=0.1$ (the contribution due to the $\textit{U}-Z-Z$
vertex and $\lambda_0=1$).} \label{Ztaumudu}
\end{figure}

\begin{figure}[htb]
\vskip -3.0truein \centering \epsfxsize=6.8in
\leavevmode\epsffile{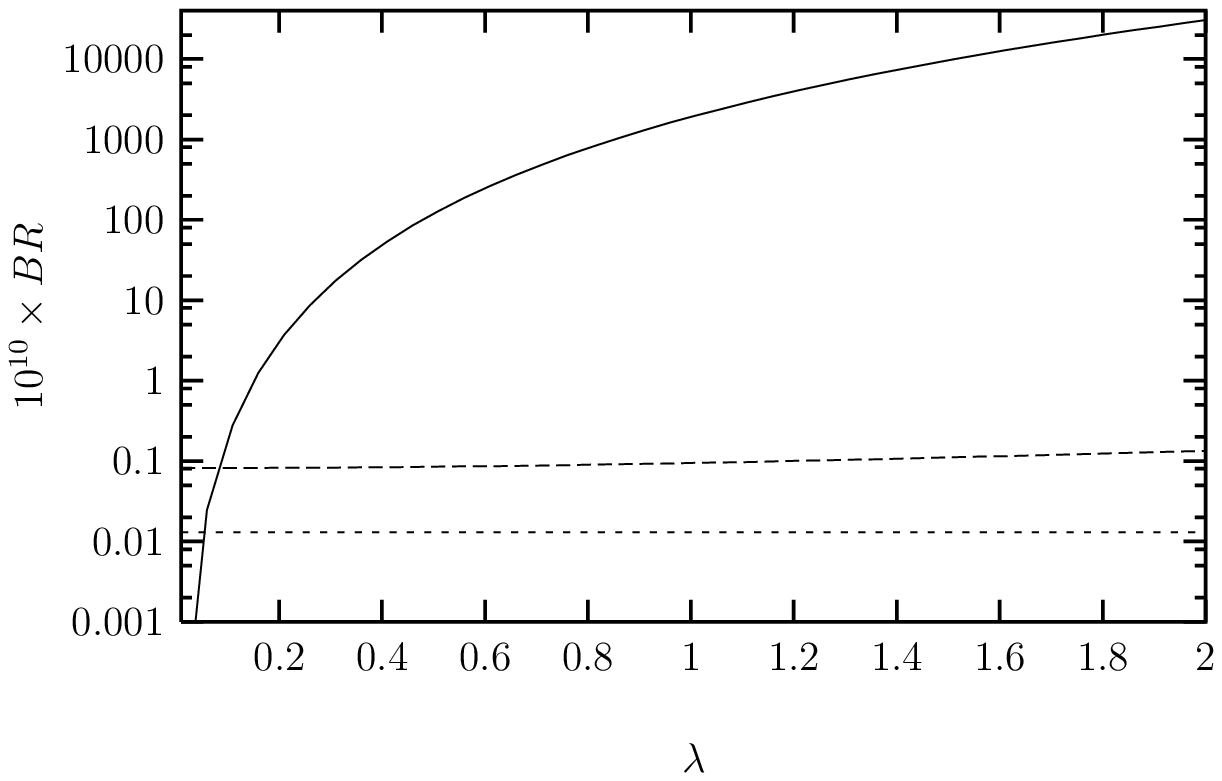} \vskip -3.0truein \caption[]{
The BR $(Z\rightarrow \tau^{\pm}\, \mu^{\pm})$ with respect to the
couplings $\lambda$, for $d_u=1.2$. Here the solid (dashed-small
dashed) line represents the BR with respect to $\lambda$ for
$\lambda=\lambda_{ee}=\lambda_{\mu\mu}=\lambda_{\tau\tau}$,
$\lambda_{ij}=0.5\, \lambda$, $\lambda_0=0.1$ and $\Lambda_u=10\,
TeV$ (with respect to $\lambda_0$ for $\lambda_{ee}=0.01$,
$\lambda_{\mu\mu}=0.1$, $\lambda_{\tau\tau}=1$,
$\lambda_{ij}=0.005$, $\Lambda_u=1.0\, TeV$- with respect to
$\lambda_0$ for $\lambda_{ee}=0.01$, $\lambda_{\mu\mu}=0.1$,
$\lambda_{\tau\tau}=1$, $\lambda_{ij}=0.005$, $\Lambda_u=10\,
TeV$).} \label{Ztaumulam}
\end{figure}

Fig.\ref{Ztaumudu} devotes the BR $(Z\rightarrow \tau^{\pm}\,
\mu^{\pm})$\footnote{For the BR $(Z\rightarrow \tau^{\pm}\,
e^{\pm})$ decay we get almost the same results and we do not
present the corresponding figures.} with respect to the scale
parameter $d_u$, for the energy scale $\Lambda_u=10\, TeV$, the
couplings $\lambda_{ee}=0.01$, $\lambda_{\mu\mu}=0.1$,
$\lambda_{\tau\tau}=1$ and $\lambda_{ij}=0.005$, $i\neq j$. Here
the solid (dashed) line represents the BR  for total contribution
and $\lambda_0=0.1$ (the contribution due to the $\textit{U}-Z-Z$
vertex and $\lambda_0=1$). The BR enhances up to the values of the
order of $10^{-8}$, for $d_u<1.1$ and the increasing values of the
scaling dimension $d_u$ results in the considerable suppression in
the BR. The contribution due to the $\textit{U}-Z-Z$ vertex is
more than two orders smaller than the total one for $\lambda_0=1$
and it shows that the effect of the $\textit{U}-Z-Z$ vertex
becomes weaker for heavy flavor outputs. In Fig.\ref{Ztaumulam},
we present the BR $(Z\rightarrow \tau^{\pm}\, \mu^{\pm})$ with
respect to the couplings $\lambda$, for $d_u=1.2$. Here the solid
(dashed-small dashed) line represents the BR with respect to
$\lambda$, for
$\lambda=\lambda_{ee}=\lambda_{\mu\mu}=\lambda_{\tau\tau}$,
$\lambda_{ij}=0.5\, \lambda$, $\lambda_0=0.1$ and $\Lambda_u=10\,
TeV$ (with respect to $\lambda_0$ for $\lambda_{ee}=0.01$,
$\lambda_{\mu\mu}=0.1$, $\lambda_{\tau\tau}=1$,
$\lambda_{ij}=0.005$, $\Lambda_u=1.0\, TeV$- with respect to
$\lambda_0$ for $\lambda_{ee}=0.01$, $\lambda_{\mu\mu}=0.1$,
$\lambda_{\tau\tau}=1$, $\lambda_{ij}=0.005$, $\Lambda_u=10\,
TeV$). For the flavor blind diagonal (off diagonal) couplings, the
BR can reach to the values of the order of $10^{-6}$ for
$\lambda=1$.  On the other hand the BR is not sensitive to the
coupling $\lambda_0$ and, for the energy $\Lambda_u=1.0\, TeV$,
its numerical value is almost one order greater compared to the
one for $\Lambda_u=10\, TeV$.

As a summary, the LFV Z boson decays are strongly sensitive to the
unparticle scaling dimension $d_u$ and, for its small values $d_u
< 1.1$, there is a considerable enhancement in the BR. In the case
that the diagonal (off diagonal) couplings are flavor blind and of
the order of one, the BR can reach to the values of the order of
$10^{-6}$ ($10^{-7}$) for the decay $Z\rightarrow \tau^{\pm}\,
l^{\pm}$, $l=\mu$ or $e$ ($Z\rightarrow \mu^{\pm}\, e^{\pm}$).
With the forthcoming more accurate measurements of the decays
under consideration it would be possible to test the possible
signals coming from the new physics which drives the flavor
violation, here is the unparticle physics.
\newpage
\newpage
\newpage
\end{document}